%% file: main_june2020.tex
\documentclass[12pt,authoryear]{elsarticle}
\usepackage[margin=1in]{geometry}
\usepackage{float}
\usepackage{amsmath,amssymb,amsthm,mathtools,amsfonts,bm,setspace,graphics,graphicx,url,caption,multicol,longtable,lscape,verbatim,color,lineno,gensymb,booktabs,hyperref,natbib,lineno,natbib}
\usepackage{subcaption}
\usepackage{algorithm}
\usepackage{algorithmic}
\DeclarePairedDelimiter{\ceil}{\lceil}{\rceil}
\usepackage[capitalise]{cleveref}
 \usepackage{setspace, subfiles} 

\usepackage{xr}
\makeatletter
\newcommand*{\addFileDependency}[1]{
  \typeout{(#1)}
  \@addtofilelist{#1}
  \IfFileExists{#1}{}{\typeout{No file #1.}}
}
\makeatother

\newcommand*{\myexternaldocument}[1]{%
    \externaldocument{#1}%
    \addFileDependency{#1.tex}%
    \addFileDependency{#1.aux}%
}
\myexternaldocument{WebSupplement}
\usepackage[english]{babel}
\newcommand{\bA}{\mathbf{A}}
\newcommand{\bx}{\mathbf{x}}
\newcommand{\bX}{\mathbf{X}}
\newcommand{\bE}{\mathbf{E}}
\newcommand{\bs}{\mathbf{s}}
\newcommand{\bn}{\mathbf{n}}
\newcommand{\bS}{\mathbf{S}}
\newcommand{\bM}{\mathbf{M}}
\newcommand{\bN}{\mathbf{N}}
\newcommand{\tR}{\mathbb{R}}
\newcommand{\bY}{\mathbf{Y}}
\newcommand{\bW}{\mathbf{W}}
\newcommand{\bG}{\mathbf{G}}

\newcommand{\bI}{\mathbf{I}}

\begin{document}

\begin{frontmatter}
\title{Group linear non-Gaussian component analysis with applications to neuroimaging}


\author[cornell]{Yuxuan Zhao}
\author[cornell]{David S. Matteson}
\author[kki,jhu1]{Mary Beth Nebel}
\author[kki,jhu1,jhu2]{Stewart H. Mostofsky}
\author[emory]{Benjamin Risk}
\ead{brisk@emory.edu}

\address[cornell]{Department of Statistics and Data Science, Cornell University}
\address[kki]{Center for Neurodevelopmental and Imaging Research, Kennedy Krieger Institute}
\address[jhu1]{Department of Neurology, Johns Hopkins University School of Medicine}
\address[jhu2]{Department of Psychiatry and Behavioral Science, Johns Hopkins University School of Medicine}
\address[emory]{Department of Biostatistics and Bioinformatics, Rollins School of Public Health, Emory University}

\begin{abstract}
Independent component analysis (ICA) is an unsupervised learning method popular in functional magnetic resonance imaging (fMRI). Group ICA has been used to search for biomarkers in neurological disorders including autism spectrum disorder and dementia. However, current methods use a principal component analysis (PCA) step that may remove low-variance features. Linear non-Gaussian component analysis (LNGCA) enables simultaneous dimension reduction and feature estimation including low-variance features in single-subject fMRI. We present a group LNGCA model to extract group components shared by more than one subject and subject-specific components. To determine the total number of components in each subject, we propose a parametric resampling test that samples spatially correlated Gaussian noise to match the spatial dependence observed in data. In simulations, our estimated group components achieve higher accuracy compared to group ICA. We apply our method to a resting-state fMRI study on autism spectrum disorder in 342 children (252 typically developing, 90 with autism), where the group signals include resting-state networks. We find examples of group components that appear to exhibit different levels of temporal engagement in autism versus typically developing children, as revealed using group LNGCA. This novel approach to matrix decomposition is a promising direction for feature detection in neuroimaging.
\end{abstract}

\end{frontmatter}

\doublespacing
\section{Introduction}
\label{s:intro}
Independent component analysis (ICA) is a popular unsupervised learning method to identify brain networks in functional magnetic resonance imaging (fMRI) studies \citep{beckmann2012modelling}.
In fMRI experiments, the observed fMRI data represent a combination of neural activity and nuisance-related variation across many different source signals. 
Assuming that latent signals are statistically independent and have non-Gaussian distributions, ICA linearly decomposes the observed fMRI data into independent spatial maps and corresponding time courses. The maps related to neural activity are commonly called resting-state ``networks.'' 
Previous studies have used group ICA to examine differential levels of intrinsic engagement between neurotypical and atypical individuals, for example, in schizophrenia \citep{Calhoun2009FunctionalReview, du2017identifying, du2019neuromark}. Group ICA of fMRI relies upon a PCA step that may discard low variance features. In ICA, a low variance feature is a spatial component that has a time course with small variance. Low variance features contained in the ``noise'' subspace in PCA may be related to neural activity, and in particular may exhibit differential levels of engagement in neurological disorders. In contrast to ICA with PCA, linear non-Gaussian component analysis (LNGCA) can extract low-variance features and has been successfully applied to single-subject fMRI data \citep{risk2019linear}. However, most fMRI studies involve data from multiple subjects and it is unclear how to extend LNGCA to such setting.

Extending non-Gaussian matrix decomposition methods to group analyses has been challenging because considerable between-subject variability exists in both the spatial configuration and the temporal engagement of the estimated functional brain networks.
Some approaches to ICA-based group inference model between-subject variability in the temporal domain while assuming spatial signals are equal across all subjects \citep{calhoun2001method,guo2011general,eloyan2013likelihood}. Models allowing subject-specific spatial deviations from the group components have been proposed \citep{beckmann2009group, guo2013hierarchical, mejia2019template, du2013group}, but in practice, all of these options are preceded by a dimension reduction step using PCA to alleviate computational demands. 
Hereafter, when we say group ICA, we are referring to this process of performing PCA prior to estimating independent components for each subject. 

PCA processing prior to ICA (PCA+ICA) can be problematic because principal components are ranked in terms of variance explained during dimension reduction.   Hence, PCA+ICA may discard important low-variance spatial signals. As an alternative, 
LNGCA can simultaneously perform dimension reduction and extract latent signals by non-Gaussianity \citep{risk2019linear}. Because components are ranked in terms of how non-Gaussian they are instead of by how much variance they explain during dimension reduction, LNGCA is able to  recover low variance non-Gaussian signals which would otherwise be discarded by PCA+ICA.

We propose group linear non-Gaussian component analysis (group LNGCA) to extract spatial signals from fMRI data that are common across subjects, as well as signals unique to each subject. 
The proposed model is an extension of LNGCA to multi-subject data, comprising two stages where the first stage applies LNGCA to each subject to extract the top non-Gaussian signals, and the second stage decomposes each subject's non-Gaussian signals into group signals and individual signals. 
Compared with current group ICA methods, the proposed group LNGCA model has the following advantages.
(1) The proposed model can recover low-variance group signals that may contain important biological structure by using subject-level LNGCA for dimension reduction instead of PCA. 
(2) The proposed model allows subject-specific spatial deviations by decomposing non-Gaussian components into group and individual signals at the second stage.
It is hypothesized some of the individual components will correspond to artifacts that are unique to each subject, such as motion artifacts that have subject-specific spatial features. These individual components can also capture subject-specific deviations from group components. 
We also propose a new approach to estimate the number of non-Gaussian signals in each subject that accounts for the spatial correlation existing in fMRI data. 
Existing approaches to dimension estimation either assume a Gaussian model \citep{li2007estimating,beckmann2004probabilistic,minka2001automatic} or ignore the spatial correlation inherent to fMRI data \citep{nordhausen2017asymptotic, jin2017optimization}.
Through simulation studies, we show that ignoring spatial correlation led to overestimates of the number of signals, while the proposed method achieved more accurate estimation.

 

In Section \ref{sec:methods}, we introduce the proposed group LNGCA model, its estimation mechanism and the proposed test of the number of non-Gaussian signals. In Section \ref{sec:sim}, we use simulations to assess model performance in terms of group components extraction and non-Gaussian signal subspace dimension estimation. In Section \ref{sec:realdata}, we use group LNGCA to estimate components from resting-state fMRI data collected from 342 school age children, including 90 with autism and 252 typically developing, and we compare our results with those obtained from group ICA estimated using GIFT software.

\section{Methods}
\label{sec:methods}

\subsection{Group non-Gaussian component analysis model}
We propose a group LNGCA model for multi-subject fMRI data, which decomposes non-Gaussian (NG) signals into group signals and individual signals for each subject. Let $i=1,\ldots,k$ index subjects, $t=1,\ldots,T$ index time points, and $v=1,\ldots,V$ index voxels (volumetric pixels). Let $\bx_i(v)\in \mathbb{R}^T$ be a data vector of observed fMRI data from subject $i$ at voxel $v$. 

\subsubsection{Subject-level LNGCA model}
To extract group signals, we first decompose observations into orthogonal NG subspace and Gaussian subspace for each subject. Specifically, 
the LNGCA model decomposes observation $\bx_i(v)$ as
\begin{equation}
    \bx_i(v) = \bM^{\bs}_i\bs_i(v)+\bM^{\bn}_i\bn_i(v), \mbox{ for }v=1,\ldots,V,
    \label{model:LNGCA}
\end{equation}
where $\bs_i(v)\in \mathbb{R}^{q_i}$ is a vector of mutually independent NG signals with $1\leq q_i\leq T$, and  $\bn_i(v)\in \mathbb{R}^{T-q_i}$ is a Gaussian noise vector. The number of NG signals $q_i$ may vary across $i$. Mixing matrices $\bM^{\bs}_i\in \mathbb{R}^{T\times q_i}$, $\bM^{\bn}_i\in \mathbb{R}^{T\times (T-q_i)}$ satisfy that $[\bM^{\bs}_i,\bM^{\bn}_i]$ has full rank for any $i$.  Here, $\{\bx_i(v)\}_{v=1,\ldots,V}$ are observed while $\{\bs_i(v)\}_{v=1,\ldots,V}$ and $\{\bn_i(v)\}_{v=1,\ldots,V}$ are latent. 
We assume $\mathrm{E}\,\bs_i(v) = \mathbf{0}$ with $\mathrm{E}\,\bs_i(v) \bs_i'(v) = \bI$.
Additionally, assume $\mathrm{E}\,\bn_i(v) = 0$  such that $\mathrm{E}\,\bx_i(v) = 0$ and $\bn_i(v)$ has unit variance for identifiability. In practice, data are centered by their sample mean. We call $\bs_i(v)$ NG signals and $\bn_i(v)$ Gaussian noise.

PCA+ICA projects observations into a smaller dimensional subspace spanned by the top principal components and extract top NG signals from that subspace.
In contrast,
subject-level LNGCA directly projects observations into a smaller subspace spanned by the top NG signals,
which are ordered by a measure of non-Gaussianity. Thus LNGCA can capture NG signals that have small variance that may be discarded in the PCA step in PCA+ICA.
Notice the key difference between LNGCA and PCA+ICA is how they search for a low rank space, and they are equivalent if there is no dimension reduction. 

\subsubsection{Group-level LNGCA model} Assume there is at least one group signal. We further decompose $\bs_i(v)$ into orthogonal group signals $\bs_{g}(v)$ and individual signals $\bs_{I,i}(v)$, 
\begin{equation}
    \bM^{\bs}_i\bs_i(v) =\bM^g_{i}\bs_g(v)  + \bM^I_{i}\bs_{I,i}(v),  \mbox{ for }v=1,\ldots,V.
    \label{model:decompIC}
\end{equation}
where $\bs_g(v)\in \mathbb{R}^{q_g}$ is a vector of mutually independent group signals with $1\leq q_g\leq \min_iq_i$, $\bs_{I,i}(v)\in \mathbb{R}^{q_{I,i}}$ is a vector of mutually independent individual signals with $q_{I,i}=q_i-q_g$. Group signals $\bs_{g}(v)$ are shared across all subjects. 
Combining \eqref{model:LNGCA} and \eqref{model:decompIC}, we have the complete decomposition, which we call the group linear non-Gaussian component analysis model (group LNGCA):
\begin{equation}
  \bx_i(v) = \bM^g_i\bs_g(v)  + \bM^I_{i}\bs_{I,i}(v) + \bM^{\bn}_{i}\bn_i(v).
  \label{model:groupNGCA}
\end{equation}


In subject-level LNGCA, the NG signals and the mixing matrix in \eqref{model:LNGCA}, 
i.e. $\bs_i(v) = [\bs_g(v)', \bs_{I,i}(v)']'$ and $\bM^{\bs}_{i} = [\bM^g_{i},\bM^I_{i}]$,
are identifiable up to sign and permutation \citep{risk2019linear}. 
Consider the matrix of subject-level components for the $i$th subject: $\bS_i \in \tR^{q_i \times V}$, with rows $[s_{ij}(1),\dots,s_{ij}(V)]$ for $j=1,\dots,q_i$. Then the group components are matched rows from different subjects, and individual components are rows that are not equal. 
In practice, we will use a singular value decomposition of the concatenated subject-level NG components (subspaces of $\tR^V$) to find the group subspace, where the size of the singular values characterizes the extent to which a direction of the non-Gaussian subspace is present across subjects; this is discussed in the next section. In applications, if a group component is prominent in one subgroup but reduced in another subgroup, then this can be captured with a group component in which the variance of the subject-specific time courses (columns of $\bM^g_{i}$) reflect differing roles, as explored in Section \ref{sec:realdata}. We expect that subject-specific components will arise from 1) components that are unique to a subject, such as certain types of motion artifacts, and 2) spatial patterns of common resting-state networks that are unique to a subject. For example, in fMRI, different subjects may have similar, but not identical, default mode networks (DMNs). Roughly speaking, the intersection of the resting-state networks corresponds to a group component, while subject-specific deviations can be allocated to the individual components.

\subsection{Estimation}
In the population formulation of group LNGCA, the decomposition for each subject following \eqref{model:groupNGCA} is a re-labeling of the subject components in \eqref{model:LNGCA} into group and individual NG components. In a sample of observations, the subject-specific LNGCAs will result in noisy estimates of the population signals, and we propose a scalable algorithm to extract the group components. We summarize our estimation procedure in Algorithm \ref{alg:groupNGCAEst}. 

\begin{algorithm}
\textbf{Input:} pre-whitened $\bX_i \in \mathbb{R}^{T\times V}$ with $\bX_i \bm{1} = \bm{0}$ and $\frac{1}{V}\bX_i \bX_i' = \bI$, NG subspace dimension $q_i$ for each subject $i$; group NG subspace dimension $q_g$.  \\
\textbf{Output:} Group signals $\hat \bS_{g}\in \mathbb{R}^{q_g\times V}$; individual signals $\hat \bS_{I,i}\in \mathbb{R}^{(q_i - q_g)\times V}$ for each subject $i$.
\begin{enumerate}
            \item Perform the subject-level LNGCA for each $i$: 
 \[     \bX_i = \hat\bM_i\hat\bS_i  + \hat\bE_i  \]
 where $\hat\bS_i \in \tR^{q_i \times V}$.
            \item Concatenate $\hat\bS_i$: $\hat\bS_{1:k} = [\hat\bS_1',\dots,\hat\bS_k']' \in \mathbb{R}^{(\sum_iq_i)\times V}$. Conduct an SVD of $\hat\bS_{1:k}$. Retain the first $q_g$ right singular vectors of $\hat\bS_{1:k}$, denoted as $\hat\bY_g \in \mathbb{R}^{q_g\times V}$.
            \item Perform noise-free ICA on the group signals subspace: 
  \[\hat\bY_g = \hat\bM\hat\bS_g  \]
 where $\hat\bS_g \in \tR^{q_g \times V}$ are the estimated $q_g$ group signals satisfying $\frac{1}{V}\hat\bS_g\hat\bS_g'=\mathrm{I}$.
            \item Project into the individual non-Gaussian subspace and apply LNGCA:
 $$\hat{\bM}_{i} \hat\bS_i(\mathrm{I} - \frac{1}{V}\hat\bS_g'\hat\bS_g)  = \hat \bA_i, $$
 $$ \hat \bA_i = \hat{\bM}^I_{i} \hat{\bS}_{I,i}$$
  where $\hat\bS_{I,i} \in \tR^{(q_g-q_i) \times V}$ is the estimated $(q_g-q_i)$ individual signals for subject $i$.
\end{enumerate}

\caption{group LNGCA Estimation Algorithm}
\label{alg:groupNGCAEst}
\end{algorithm}

In Step 1, we estimate NG signals for each subject. In this paper, we use the logistic non-linearity with the fastICA algorithm \citep{hyvarinen1999fast} modified for LNGCA 
because the logistic non-linearity is used in Infomax and performs well in fMRI \citep{correa2007performance, plis2014deep}.
We find this algorithm works well for the super-Gaussian distributions found in fMRI (see simulations in Section \ref{sec:sim}), and it is computationally tractable and less sensitive to initializations. Our approach can be adapted to more flexible (but typically computationally costly) non-linearities that also allow the estimation of sub-Gaussian densities, for example, ProDenICA \citep{hastie2002independent}, if applied to other applications with sub-Gaussian distributions. 

In Step 2, we construct the subspace spanned by all NG signals across subjects by concatenating the subject-level NGs. To estimate the group non-Gaussian subspace, we conduct an SVD and extract the first $q_g$ singular vectors, which is equivalent to performing PCA on the concatenated subject NGs. Insight into this step can be gained by viewing the SVD as a principal angle analysis. 
For ease of notation in what follows, rescale the NG components to have norm equal to 1 prior to concatenation. 
Assume observations $v=1,\dots,V$ from $k$ subjects, and let $\bS^g_{i,d} \in \tR^{1 \times V}$ be the $d$th group component in the $i$th subject. 
In the concatenated NG components, this group component has multiplicity $k$. 
Then the size of the singular value from a group component shared by all subjects is equal to $\sqrt{k}$. In practice, this group subspace will be an average of the information shared across subspaces. For two subjects, the principal angles between their subspaces are equal to $\arccos (\sigma_{d}^2-1)$, $d=1,\dots,(q_1+q_2)$, where $\sigma_d$ is the $d$th singular value and $q_i$ is the number of NG components in the $i$th subject. In the case of $q_1=q_2=1$, the first right singular vector is the average of the NG components, and $\sigma_1^2-1$ is the correlation (for mean centered data). The use of the SVD is similar to the approach in \citep{feng2018angle} in which joint structure is extracted from multiple datasets, but here we apply the procedure to the concatenated NG signals from multiple subjects to obtain a subspace that is shared across multiple subjects. 

We discuss how to estimate $\{q_i\}_{i=1,\ldots,k}$ in Section \ref{sec:dimest}. 
Determining $q_G$ is beyond the scope of this work but discussed in Section \ref{s:discuss}.
Notice the choice of $q_G$ is more of a practical consideration for large noisy fMRI data,
since those data usually do not admit a clear gap between group and individual components.
With small $q_G$, only a small portion of NG signals can be estimated, but fast and accurate.
With large $q_G$, 
the returned components are likely to contain all NG signals,
although with slow computation, and visual inspection is required to exclude noise components.
We show such phenomena through simulation study.

In Step 3, we search for the orthogonal transformation of the group subspace that maximizes non-Gaussianity using noise-free ICA. This produces an estimate of the group components. 

In Step 4, we estimate the subject-specific signal subspace as the orthogonal complement of the group signal subspace in the estimated NG subspace for that subject. Then we apply LNGCA to the subject-specific subspace to extract subject-specific NG signals.



\subsection{Comparison with group ICA model and algorithm}

A major difference between group ICA and group LNGCA is that LNGCA is conducted for subject dimension reduction rather than PCA. In simulations, we examine how this impacts the estimation under different signal variance regimes. 
Although both algorithms use PCA for the group stage dimension reduction,
it is performed on distinct subspaces.
For group LNGCA, as long as the NG signals are extracted in many subjects, 
they will be the top principal components in the concatenated NG components.
For group ICA,
there is no such guarantee.
An NG signal is included in the 
top principal components only when it has relatively large variance and thus is not largely discarded in the subject PCA step.
Although one could keep a very large ratio of variance in the subject PCA step to avoid the issue,
that would result in challenges to accurately conduct the group stage PCA (usually through iterative methods) due to much larger data size
compared to group LNGCA.
When there is no dimension reduction on the subject data, the group LNGCA algorithm is equivalent to group ICA algorithm as in \cite{calhoun2001method}.

In addition to the difference in how the group subspace is defined, group LNGCA estimates individual signals, which is useful for examining non-Gaussian signal that is not captured in the group component. In contrast, group ICA discards all information not captured by the group components.

\subsection{Test the dimension of the non-Gaussian signals subspace}
\label{sec:test}
To estimate the NG subspace dimension $q_i$ for each subject, recently proposed methods \citep{nordhausen2017asymptotic, jin2017optimization} sequentially test the dimension of the NG subspace:
\begin{align*}
    H_0^k: \mbox{There are at most } k \mbox{ NG signals.} \quad \mbox{ versus } \quad
    H_A^k: \mbox{There are at least } k+1 \mbox{ NG signals.}
\end{align*}
for $0\leq k\leq T-1$. Suppose the true dimension is $k_0$. With increasing sample size, we expect a good test to perform as: (1) for $k<k_0$, the power of the test for $H_0^k$ approaches one; (2) for $k=k_0$, the size of the test for $H_0^{k_0}$ approaches prespecified $\alpha$; and (3) for $k>k_0$, the rejection probability for $H_0^k$ tends to be smaller than $\alpha$.

Denote the p-value associated with $H_0^k$ by $p_k$. The estimate is $\hat k=\{k\;|\;p_{\hat{k}}\leq \alpha,\; p_{\hat{k}-1}>\alpha\}$. It is very expensive to test $H_0^k$ for all possible $k$. However, using a binary search in $k$, we can expect no more than $\ceil{\log_2T}$ tests for dimension $T$ \citep{jin2017optimization}. 
The estimated $\hat k$ then relies on multiple tests, which may be problematic for large $T$. One may consider adjusting the obtained p-values using Bonferroni correction or FDR control. However, the sequential tests are highly dependent in that: for any $k_1<k_2$, when $H_0^{k_1}$ holds, $H_0^{k_2}$ must hold. Adjusting p-values without considering this special dependence structure may largely decrease the power of sequential tests and damage the estimated $\hat k$. Thus we use the original p-values and show through simulations in Sec \ref{sec:dimest} that we obtain accurate estimation of $k$. 

\subsubsection{Motivation}
Suppose $D(\cdot)$ is some non-Gaussianity measure for a signal, e.g., skewness, excess kurtosis, or the Jarque-Bera statistic. 
For a matrix $\bY\in \mathbb{R}^{T\times V}$, sort $\bY = [\bY_{(1)}',\cdots, \bY_{(T)}']'$ such that $D(\bY_{(1)})>\cdots>D(\bY_{(T)})$, where $\bY_{(t)} \in \mathbb{R}^{V}$ is a row vector of $\bY$, for $t=1,\ldots,T$. In this section, we drop the subject index, but in practice, these tests are applied separately to each subject. We will use this sorting (rank) notation for other data matrices below. 
Motivated by  \cite{jin2017optimization}, we assume 
$\bY_{(j_1)}$ is more non-Gaussian than $\bY_{(j_2)}$ in terms of $D(\cdot)$ for $j_1<j_2$. Consequently, under $H_0^k$, $[\bY_{(1)}',\ldots,\bY_{(k)}']'$ are NG signals and $[\bY_{(k+1)}',\ldots, \bY_{(T)}']'$ are Gaussian noise. 
\cite{jin2017optimization} proposed an algorithm to test $H_0^k$ based on a max-min estimator which maximizes the non-Gaussanity of $k$ components while minimizing the non-Gaussianity of the $(T-k)$ Gaussian components. 
The key idea is to extract the NG signals and Gaussian noise from the original data, then generate new data samples mixed from the extracted NG signals and new sampled Gaussian noise.
However, for every generated new sample, the algorithm conducts a full rank ICA-like algorithm on a matrix of size $V\times T$, 
which is very expensive. 
To overcome such computational burden, we choose to only sample the Gaussian noise rather than the whole data set.


\subsubsection{Resampling test for dimension estimation}
We propose an algorithm based on the sample distribution of the maximum component-wise non-Gaussianity, i.e. $D(\bG_{(1)})$, when ICA is applied to $T-k$ Gaussian components $\bG$. We state our method in Algorithm \ref{alg:ourDimTestAlg}.

\begin{algorithm}
\textbf{Input:} pre-whitened $\bX \in \mathbb{R}^{T\times V}$ with $\bX \bm{1} = 0$ and $\frac{1}{V}\bX \bX' = \bI$, number of samples $B$.\\
\textbf{Output:} Empirical p-value $\hat p$.
        \begin{enumerate}
            \item Estimate $\hat \bS=\hat \bW\bX$ with $\hat \bS\in \mathbb{R}^{T\times V}, \hat \bW\in \mathcal{O}^{T\times T }$.
            \item Sort $\hat\bS = [\hat\bS_{(1)}',\cdots, \hat\bS_{(k+1)}',\ldots,\hat\bS_{(T)}']'$ to obtain $D(\hat\bS_{(k+1)})$.
            \item Repeat for $b=1,\cdots,B$:
            \begin{enumerate}
                \item Generate $(T-k)$ independent Gaussian components $\bG^{(b)}\in \mathbb{R}^{(T-k)\times V}$.
                \item Estimate $\hat\bS_G^{(b)}=\hat \bW_G^{(b)}\bG^{(b)}$ with $\hat \bS_G^{(b)}\in \mathbb{R}^{(T-k)\times V}, \hat \bW_G^{(b)}\in \mathcal{O}^{(T-k)\times (T-k)}$.
                \item Sort the rows of $\hat\bS_G^{(b)}$ to obtain $D(\hat\bS^{(b)}_{G,(1)})$.
                
            \end{enumerate}
            \item Calculate the empirical p-value: 
            $\hat{p}=\frac{\# \{D(\hat\bS_{(k+1)}) < D(\hat\bS^{(b)}_{G,(1)})\}}{B}$
        \end{enumerate}
    \caption{Our NG subspace dimension test $H_0^k$.}
    \label{alg:ourDimTestAlg}
\end{algorithm}
Now in each repetition, 
the ICA is implemented on a matrix with only $T-k$ components at \cref{alg:ourDimTestAlg} step 3(b),
to extract the component $\hat\bS^{(b)}_{G,(1)}$ with highest $D(\cdot)$ among orthogonally transformed $(T-k)$-variate Gaussian components.
We assume that $\hat\bS_{(k+1)}$ shares the same distribution as $\hat\bS^{(b)}_{G,(1)}$. As a result,  $D(\hat\bS_{(k+1)})$ and $D(\hat\bS^{(b)}_{G,(1)})$ have the same distribution. 


To implement our proposed Algorithm, we use FOBI \citep{cardoso1989source} with $D(\cdot)$ equal to kurtosis. Unlike other methods, FOBI has a closed-form solution, which makes it computationally scalable and avoids the need for multiple restarts. Although in theory other more computationally intensive methods could be used, our goal is to develop a method that can be applied to real fMRI data. Previous approaches without spatial correlation have found FOBI works well for dimensionality tests \citep{nordhausen2017asymptotic}. Note the goal here is to estimate the dimensions, rather than accurate components.


\subsubsection{Spatially correlated Gaussian noise}
For fMRI data, 
the pre-processing steps may apply Gaussian smoothing, and interpolation steps also introduce smoothness \citep{Chen2018EffectConnectivity,Coalson2018TheAreas}. For example, suppose $\bG \in \tR^{(T-k) \times V}$ has each row generated from a Gaussian random field. Then unmixing these components can result in disk-like features in spurious NG components \citep{risk2019linear}. 
As we will show, methods assuming independent entries in Gaussian noise can overestimate $k_0$,
and consequently too large $k_0$ cannot help achieve dimension reduction.
We can account for the spatial autocorrelation by adjusting Algorithm \ref{alg:ourDimTestAlg} step 3(a). 

When there is no spatial correlation, we generate row vectors $\bG^{(b)}_l\overset{i.i.d.}{\sim} \mathcal{N}(0, \mathrm{I}_V)$ for $l=1,\ldots,T-k$. When there is spatial correlation, we can modify Step 3(a) in Algorithm \ref{alg:ourDimTestAlg}. We first specify or estimate the spatial correlation matrix $\Sigma \in \mathbb{R}^{V\times V}$, then generate row vectors $\bG^{(b)}_l\overset{i.i.d.}{\sim} \mathcal{N}(0, \Sigma)$ for $l=1,\ldots,T-k$. In practice, we can generate random fields by applying smoothers to iid Gaussian data, as implemented in \texttt{neuRosim} \citep{welvaert2011neurosim}. 
Such modification ensures a component is only classified as NG signal when it is more non-Gaussian than the spatially correlated Gaussian noise.

\section{Simulations: spatio-temporal signals}
\label{sec:sim}
We evaluate the performance of group LNGCA versus group ICA \citep{calhoun2001method} in a simulation study with twenty subjects. We also compare the performance of our NG subspace dimension test with those proposed in \cite{nordhausen2017asymptotic}.
All algorithms are repeated with $30$ random initializations.
Our code is available at \url{https://github.com/yuxuanzhao2295/Group_LNGCA_for_Neuroimaging}.


\subsection{Data generation}
\begin{figure}[htbp]
\centering
\includegraphics[width=0.7\linewidth]{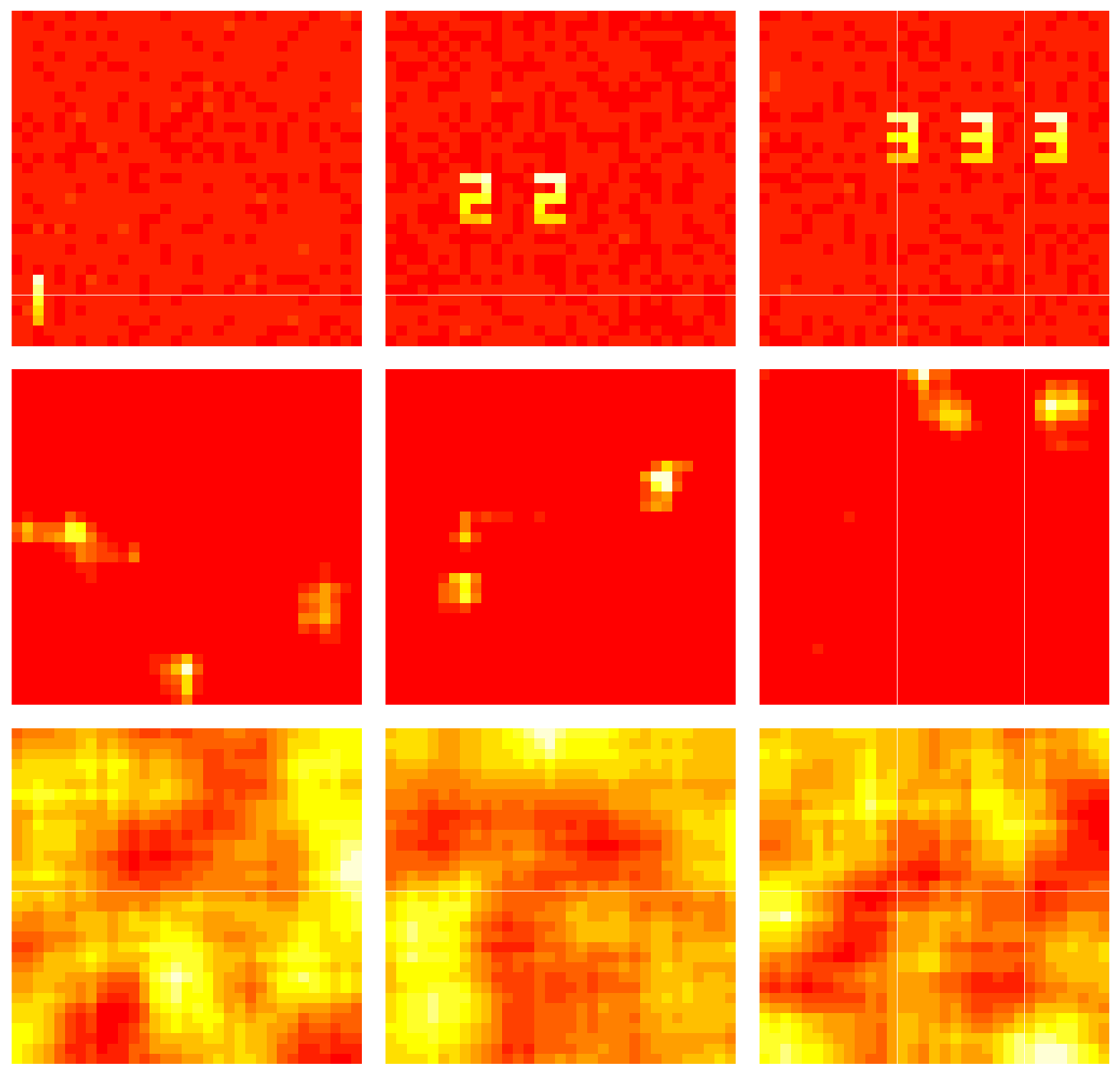}
\caption{Simulated non-Gaussian and Gaussian components. First row depicts $3$ group NG signals. Second row depicts $3$ (of $22$) individual NG signals. Last row depicts $3$ (of $25$)  Gaussian noise components. Each component is a $33\times 33$ image corresponding to $V=1089$.}
\label{fig:Signal}
\end{figure}

For each of the twenty subjects, we used $3$ group signals, $22$ individual signals from gamma random fields, and $25$ Gaussian noise components from Gaussian random fields. 
Example components are depicted in  \cref{fig:Signal}. 
The group signals have active pixels in the shape of a ``1", ``2 2", or ``3 3 3" with values between $0.5$ and $1$ and inactive pixels as  iid normal with mean $0$ and variance $0.001$.
Random fields (RF) were simulated using the R package \texttt{neuRosim}.
For both gamma and Gaussian random fields, the full-width at half maximum of the Gaussian kernel, which controls spatial correlation, was set to 9. For the marginal distributions of the individual signals, we selected gamma parameters (shape parameter $0.02$, rate parameter $10^{-4}$) that resulted in individual signals with mean logistic non-linearity ranging from $-1.17$ to $-0.87$. The corresponding range found in our real data application is from $-1.39$ to $-0.82$. For the columns of the mixing matrix, we simulate AR$(1)$ processes per subject with $\phi=0.37$, estimated from our real data application (detailed in the Web Appendix \cref{sec:sim_setup}). The scaling of the columns was chosen to result in variance proportions detailed below.  

To control signal and noise strength, we define the subspace variance ratio (SVAR) among the group NG subspace, individual NG subspace, and Gaussian noise subspace. 
Fixing the subscript $i$,
let $\lambda_1,\ldots,\lambda_{q_g}$ be the nonzero eigenvalues from the eigenvalue decomposition (EVD) of the covariance matrix of $\bM_{i}^g\bS_g$; $\mu_1,\ldots,\mu_{q_i-q_g}$ be the nonzero eigenvalues from the EVD of the covariance matrix of $\bM_{i}^I\bS_{I,i}$; and $\nu_1,\ldots,\nu_{T-q_i}$ be the nonzero eigenvalues from the EVD of the covariance matrix of $\bM_{i}^{\bn}\bN_{i}$. The SVAR is defined as $\sum_{l=1}^{q_g}\lambda_l:\sum_{l=1}^{q_i-q_g}\mu_l:\sum_{l=1}^{T-q_i}\nu_l$. We can also normalize these values by the total variance, and then this is equivalent to the proportion of variance from the group signals, individual signals, and Gaussian noise. Here, we focus on the variance ratio of the group NG subspace, so we designed three simulation settings corresponding to a high SVAR (large $\sum_{l=1}^{q_g}\lambda_l$), a medium SVAR, and a low SVAR in the group NG subspace. 
In the discussion that follows, we refer to the variance of a component as the sum of squares of the corresponding column of the mixing matrix.

 First, we describe how we allocated the variance in a given subspace to the components in that subspace. For all settings, we assumed the SVAR was equal for all $20$ subjects. Then in the group subspace, we defined a low variance component, a medium variance component, and a higher variance component using the $0.1$, $0.5$, and $0.9$ quantiles of the variance of the group signals estimated in the real data application ($15.4\%:29.8\%:54.8\%$, based on 
 the 59 group components estimated from 342 subjects). This is designed to examine whether signals with lower variance may be discarded by group ICA. In the individual NG subspace and the Gaussian subspace, we assumed all components had equal variance. 

Next, we describe how we allocated the total variance to each subspace. To determine the high SVAR scenario, we calculated the proportion of variance in each of these subspaces for each subject in the real data application, wherein we estimated $59$ group signals and $85$ total NG signals 
with either $128$ or $156$ time points, see Section \ref{sec:realdata}. We then calculated the median across subjects and re-scaled the proportions to sum to one ( $\sum_{l=1}^{q_g}\lambda_l:\sum_{l=1}^{q_i-q_g}\mu_l:\sum_{l=1}^{T-q_i}\nu_l=33.5\%:29.9\%:36.6\%$). Note this distributes the variance of the NG subspace formed from 59 signals to 3 signals, and thus these signals have a proportionately very large variance. For the low SVAR scenario, we set the variance of each of the group signals  equal to the $0.1$, $0.5$, and $0.9$ quantiles of the variances of the 59 group signals. This results in a much lower variance relative to the high SVAR scenario:
$\sum_{l=1}^{q_g}\lambda_l:\sum_{l=1}^{q_i-q_g}\mu_l:\sum_{l=1}^{T-q_i}\nu_l=1.7\%:46\%:52.3\%$. For the medium SVAR scenario, we set $\sum_{l=1}^{q_g}\lambda_l$ to be half of the sum of that in high and low SVAR scenario. 
Thus we have medium SVAR: $\sum_{l=1}^{q_g}\lambda_l:\sum_{l=1}^{q_i-q_g}\mu_l:\sum_{l=1}^{T-q_i}\nu_l=17.6\%:38.6\%:43.8\%$.

\subsection{Non-Gaussian subspace dimension estimation}
\label{sec:dimest}
\begin{figure}
    \centering
    \includegraphics[width=0.9\linewidth]{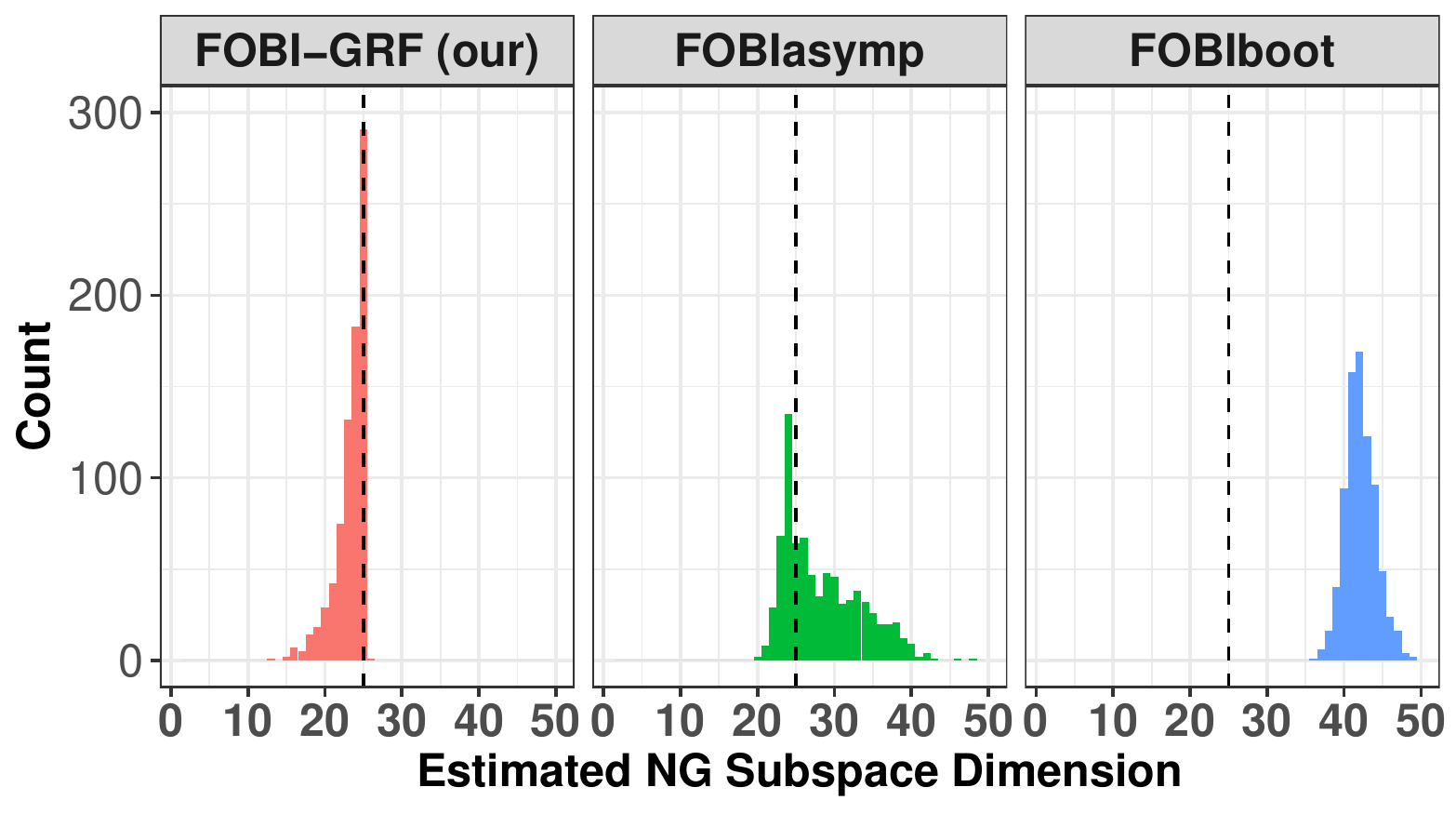}
    \caption{Estimated non-Gaussian subspace dimension across $800$ subjects under medium SVAR setting. The significance level $\alpha=0.05$. Dashed line indicates true dimension $25$. The most frequently selected dimension using our test, FOBI-GRF, corresponds to the true dimension. Although the test underestimated the dimensions in many simulations, this was due to possibly missing individual components, while it always retained the  group components, as described in Section \ref{sec:sim_groupcomponents}.}
    \label{fig:testRej}
\end{figure}

We implemented our test in \cref{alg:ourDimTestAlg} with spatially correlated Gaussian noise,
denoted as \texttt{FOBI-GRF},
and compare its performance with tests \texttt{FOBIasymp} and \texttt{FOBIboot} from the R package \texttt{ICtest} \citep{nordhausen2017ictest}. For each test, we apply a binary search to estimate the NG subspace dimension. For each SVAR setting, one experiment contains $20$ subjects and the experiment is repeated $40$ times, thus $800$ subjects in total. The true NG subspace dimension is $25$ for all settings. 
We report the results under the medium SVAR setting in \cref{fig:testRej}. 
The results under the high and low SVAR settings are very similar and appear in Web Appendix \cref{fig:dimest_high_low}.

\texttt{FOBIasymp} and \texttt{FOBIboot} do not consider the spatial correlation, and they clearly overestimate the dimension. \texttt{FOBIboot} is a parametric bootstrap method that simulates Gaussian noise with iid entries. We see \texttt{FOBIboot} estimates almost all components as NG when the assumption is violated. Our test \texttt{FOBI-GRF} performs the best in terms of accuracy and stability. Sometimes our test underestimates the number of NG components, but as we will show in the next section, group LNGCA is still able to extract all group NG components. The underestimation can occur due to the event of ``unmixed'' Gaussian RF components having higher non-Gaussianity than gamma RF components. This is similar to the gap between signal and noise eigenvalues in principal component analysis \citep{ProfileLik2006Zhu}. In our context, a smaller gap between the NG signals and transformations of the spatially correlated Gaussian signals can result in a decrease in the accuracy of our testing procedure. In applications, the individual components tend to correspond to structured artifacts or individual resting-state networks, which may be more distinct from Gaussian RFs than the gamma RFs used in our simulations. The choice of gamma RFs here is to allow computational scalability, since it is difficult to generate 22*20 individual NG components. 

We comment here the NG subspace dimension is a challenging problem, and in particular is more difficult for higher dimensions. Suppose we have a fixed number of NG signals and linearly mix them with Gaussian components. As the number of Gaussian components increase, there will also be more components displaying high non-Gaussianity among extracted independent components. \citet{bickel2018projection} showed the projection of multivariate Gaussian distribution can approximate any non-Gaussian distribution for large $T$, in particular, for $T>V$. Here, it appears that with low frequency some components estimated from Gaussian RFs can be more non-Gaussian than the NGs. Thus when estimating the NG subspace dimension $k$, one could consider using a higher value of $\alpha$ with increased dimension $T$, as the NG component can be hidden between spurious components. In practice, a visual inspection may suffice to identify the spurious, disc-like components that arise from maximizing non-Gaussianity of Gaussian RFs.

 \subsection{Group components extraction}\label{sec:sim_groupcomponents}
\begin{figure}
    \centering
    \includegraphics[width=0.9\linewidth]{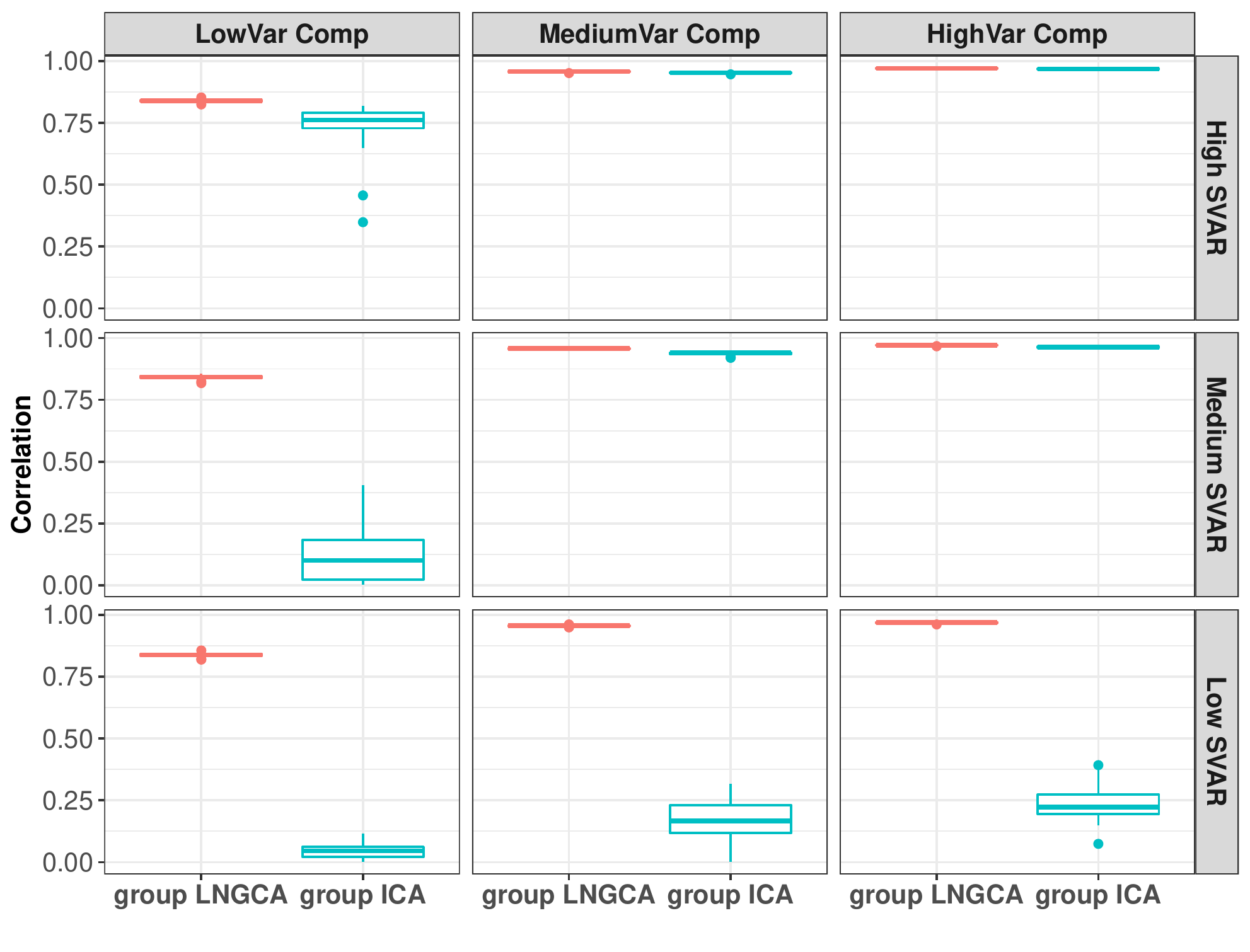}
    \caption{Correlation between three estimated and true group components. The correlation under all settings for group LNGCA concentrates at a high correlation value with vanishing variance.}
    \label{fig:corrComp}
\end{figure}

\begin{figure}
    \centering
    \includegraphics[width=0.9\linewidth, height=0.8\linewidth]{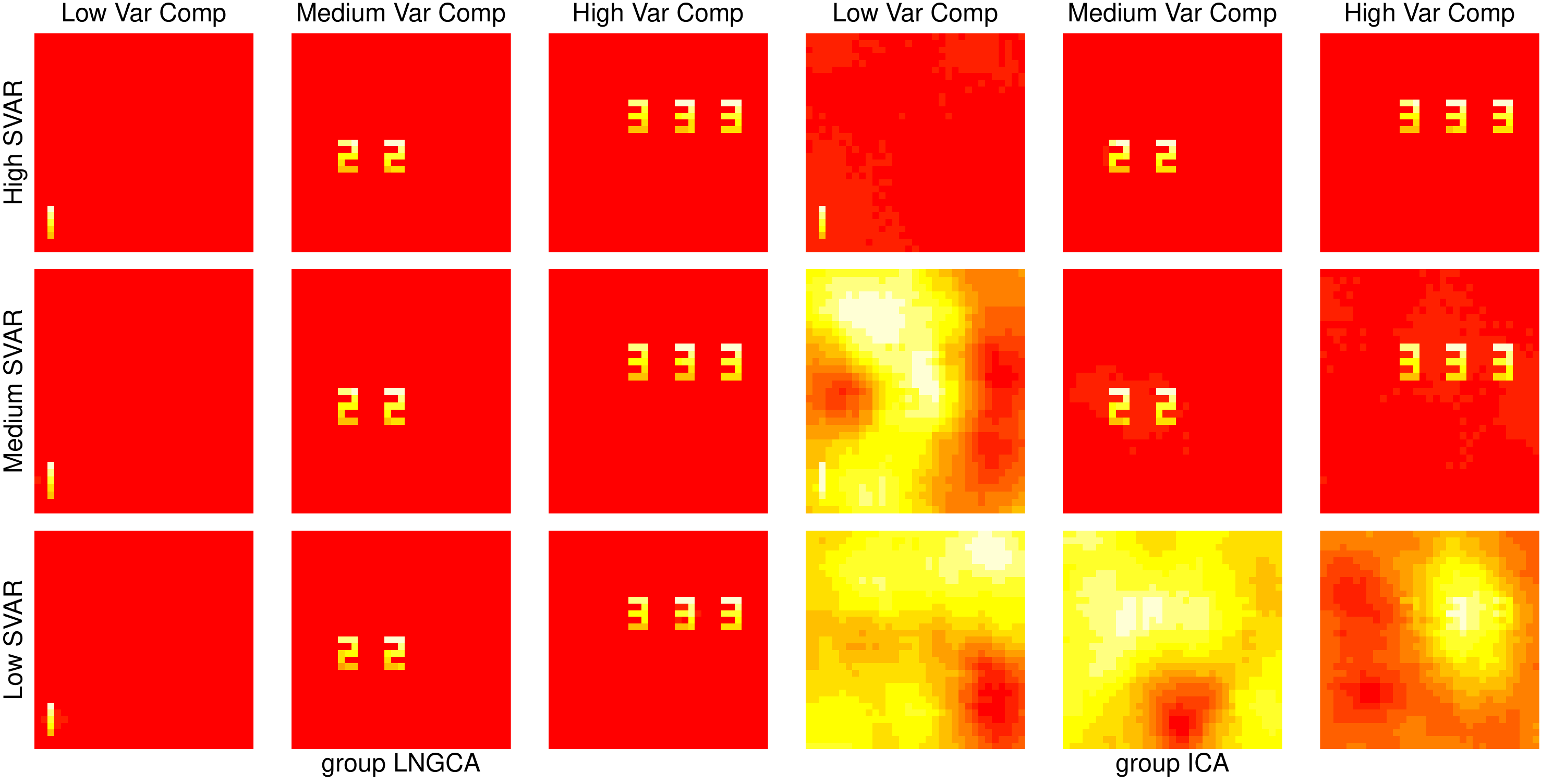}
    \caption{The estimated group components from a representative simulation (median matching error) when $q_G=3$. Left three columns display results from group LNGCA, while right three columns display results from group ICA. For each method, the allocated variance increases from left to right among three signals. Three rows represents high, medium and low SVAR settings respectively, from top to bottom.}
    \label{fig:estComp}
\end{figure}

Using the estimated subject NG subspace dimension from our test as reported in Sec \ref{sec:dimest}, we implement group LNGCA as in Algorithm \ref{alg:groupNGCAEst} to estimate the group signals. For group ICA, we conducted the subject-level PCA with the number of signals selected to retain at least $82\%$ of the variance (selected to be consistent with the real data application, detailed in Section \ref{sec:realdata}) for each subject. 
For both group LNGCA and group ICA, 
we examine their performance with group NG subspace dimension $q_g=2, 3$ and $4$.
After extracting group signals, we matched them to true signals using a modification of the Hungarian algorithm \citep{risk2014evaluation}, and then calculated the correlation between matched signals, for each method respectively. 
The simulation is repeated $40$ times for each setting. 
We first report the results when $q_g=3$:
the correlation between matched components in \cref{fig:corrComp}, and the estimated signals from the repetition associated with the median matching error (with true signal) in each setting
are depicted in \cref{fig:estComp}.  

When SVAR is high, both methods can recover three group signals with high accuracy. However, accuracy across methods diverges as SVAR decreases. Group ICA failed to recover the low variance group signal in the medium SVAR setting and failed to extract any group signals under the low SVAR setting, while group LNGCA was highly accurate under all three SVAR settings. Also notice group ICA is better able to estimate group signals with larger variance, which is consistent with the fact it reduces the data dimension by signals' variance and may lose low variance signals.

We show in the appendix that group LNGCA is robust to the specified number of group components: (1) when $q_G=4$, all three group signals can still be recovered with high accuracy; (2) when $q_G=2$, two of the three group signals can be recovered with high accuracy.


\section{Resting-state fMRI data example}
\label{sec:realdata}
We applied group LNGCA and group ICA to resting-state fMRI data from 342 school-age children, ages 8 through 12 years, recruited at the Kennedy Krieger Institue (PI: S. Mostofsky) including 90 with autism spectrum disorder (ASD) and 252 typically developing (TD) controls. Resting-state fMRI was acquired during either a 5 min 20 s- or 6 min 30 s-long scan on a 3.0 T Philips scanner using a single-shot, partially parallel, gradient-recalled echo planar sequence (repetition time 2500ms, echo time 30ms, flip angle = $75^{\circ}$, SENSE factor of 2, 3-mm axial slices with no slice gap, in-plane resolution of $3.05\times  3.15$ mm resulting in $84\times 81 \times 47$ voxels). Data were registered to MNI space and smoothed using a 6-mm FWHM Gaussian filter. Each participant's preprocessed data were mean-centered and variance normalized on a voxel-wise basis. Additional details on preprocessing and motion exclusion criteria are provided in the Web Appendix \cref{sec:real_data_preprocessing}.

\subsection{Subject-level and group-level dimension reduction}
We applied our test in \cref{alg:ourDimTestAlg} to six participants. Gaussian RFs were generated using the estimated FWHM of the data from 3dFWHMx in AFNI. 
The estimated dimension was 65, 38, 89, 47, 73 and 83, 
corresponding to participants with 128 time points for first three participants and 156 for last three participants (detailed in the Web Appendix \cref{sec:real_data_dim_est}). Thus the number of components varied by participant, but in general indicated the NG signal was contained in a lower-dimensional subspace. 
The test is applied to only six participants because it is computationally intensive on fMRI data,
and we only need to get a ballpark number here.

For the subject-level PCA in group ICA, we performed dimension estimation on each participant's preprocessed data using an information theoretic approach implemented within GIFT \citep{li2007estimating}, which calculates minimum description length (MDL) assuming a Gaussian model. The maximum of the estimated number of PCs from the 342 datasets was 59. Following recommendations from \cite{erhardt2011comparison}, we conservatively chose to use 85 components for all subjects.  This kept at least $82\% $ of the variance in each subject's data.  We note that this approach retains more variance than suggested using GIFT's dimensionality tests, and in this respect will make our results with group LNGCA more similar to group ICA.
For group LNGCA, we also chose 85 NGs for all participants, to make fair comparison between two methods. Notice $85$ is also near the maximum of our estimated NG dimension from the six participants. 

For the group-level PCA,
we retain 59 components for both group LNGCA and GIFT to make fair comparison.
As mentioned above,
59 is the maximum estimated number of PCs across all subjects.

\subsection{Group ICA and Group LNGCA}
For group ICA, 
participant-specific PCs were temporally concatenated and a second PCA was performed to extract group-level PCs.
Noise-free ICA was repeated on the group-level PCs $100$ times using the Infomax algorithm \citep{bell1995information} and the ICASSO toolbox \citep{himberg2004validating} with randomized initial conditions in GIFT. 
Group components from GIFT were labeled using ICs from \citep{allen2011baseline}.


For group LNGCA,
we used $40$ randomized initializations for each subject-level LNGCA and $100$ in the group-level ICA. We used the logistic (Infomax) non-linearity and matched the components in GIFT and group LNGCA using the Hungarian algorithm. The majority of components were very similar, with correlations greater than $0.8$; twenty components had correlations less than $0.8$, and of those twenty, five had correlations less than $0.5$.

\begin{figure}[htbp]
    \centering
    \includegraphics[width=0.7\linewidth, height=0.7\linewidth]{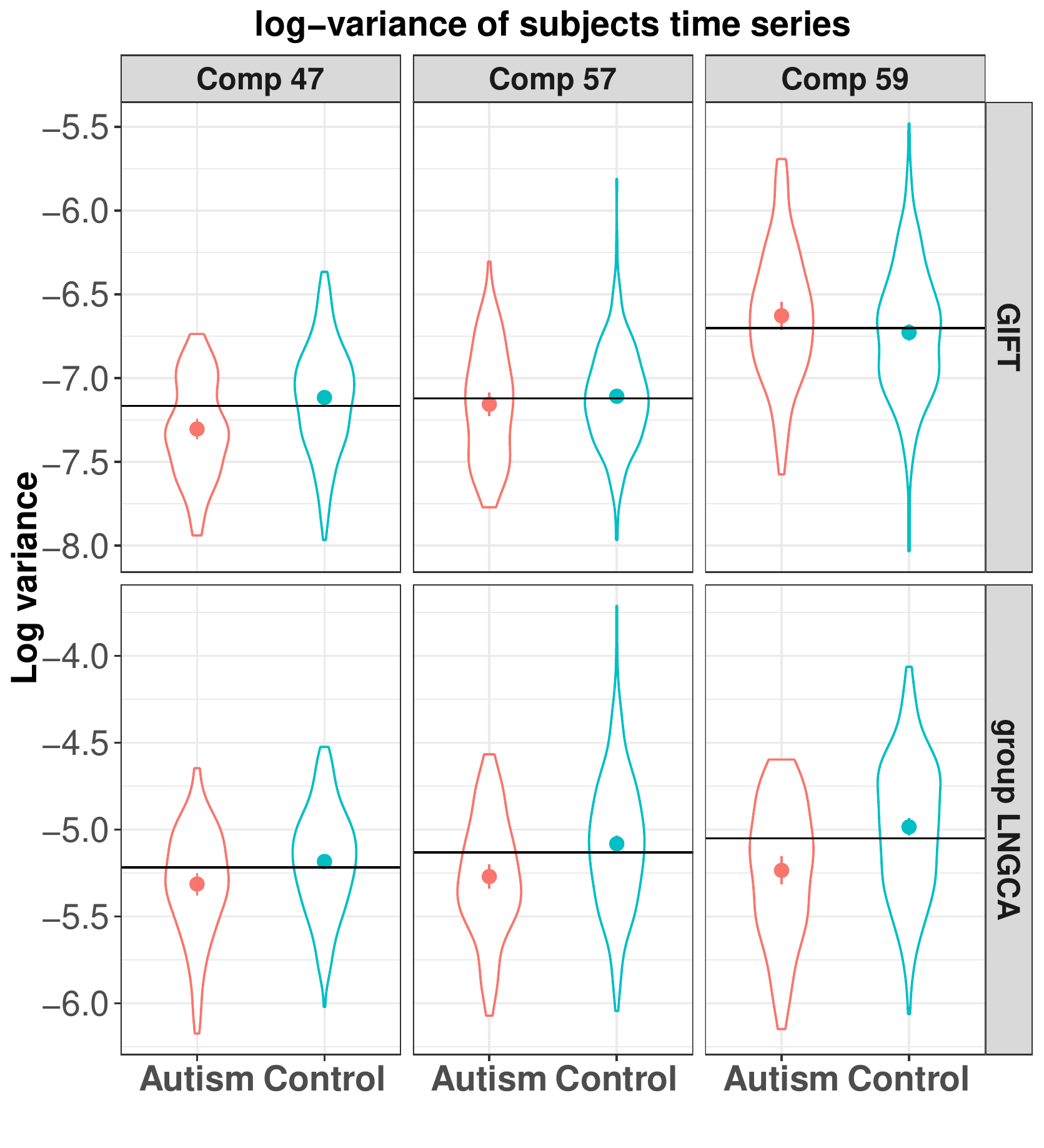}
    \caption{Most significant component: group LNGCA component 59 (p=$9e-5$, corrected), its matched GIFT component 59 (p=$0.1$) is not significant; 2nd significant component: GIFT component 47 (p=$2e-4$), its matched group LNGCA component 47 (p=$0.01$) is also significant; 3rd significant component: group LNGCA component 57 (p=$3e-3$), its matched GIFT component 57 (p=$0.4$).  Horizontal line marks the mean across all children.}
    \label{fig:pvalue}
\end{figure}
\begin{figure}[htbp]
    \centering
    \includegraphics[width=\textwidth]{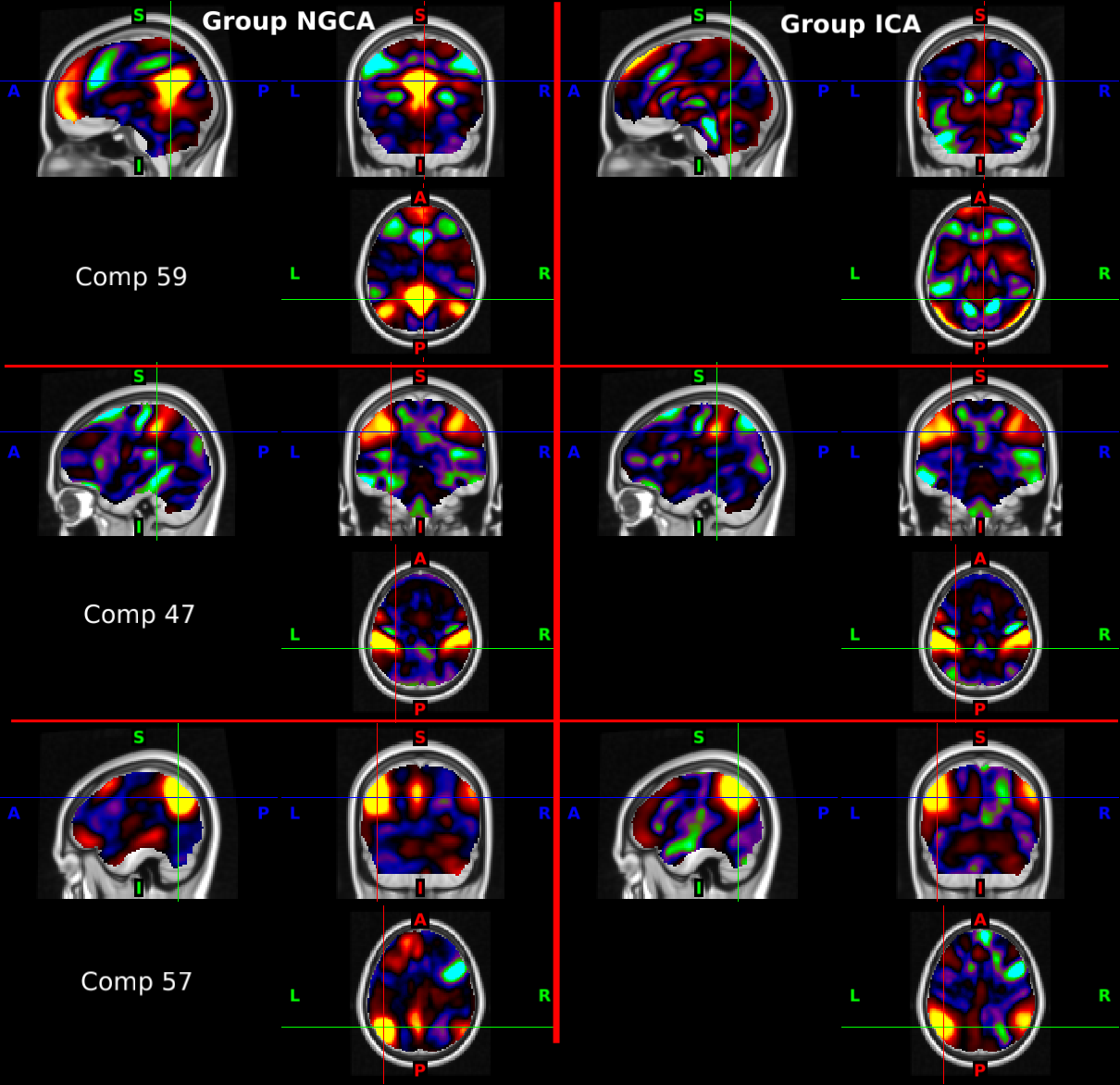}
    \caption{Comparison of group LNGCA (left) and matched group ICA components (right) associated with the most significant ASD vs. TD differences in intrinsic engagement. Color indicates voxel contribution across all subjects, and color bars are based on the second and ninety-eighth percentiles for the positive and negative values of each component. (Top row) Group LNGCA component 59 characterizes the default mode network while the matched group ICA component may be related to motion (high activation near the edges), and a significant ASD vs TD difference in engagement is only observed for the former (\cref{fig:pvalue}). For both methods, component 47 (middle row) is easily identified as the frontoparietal network and is significantly less engaged in ASD vs TD children. Component 57 (bottom row) is identified as default mode network for both methods but is more left-lateralized for group LNGCA, and we only observe a significant ASD vs TD difference in engagement for group LNGCA.
    }
    \label{fig:signals}
\end{figure}

The variance of subjects' time series corresponding to extracted group components may reveal different levels of intrinsic activity between children with autism and healthy controls. 
For a given subject,
it represents the energy that a component exhibits.
If a component is orthogonal to the subjects' data, then the variance will be zero. If a component is more intrinsically active in children with ASD compared to TD children, then we expect the variance of that component to be higher on average in the former. 
For each estimated group component (59 from group LNGCA and 59 from GIFT), we applied t-tests to the log variance of the corresponding subject time series between two groups after controlling the age and sex.
After adjusting the p-values for 118 hypothesis tests using FDR control, we found the variance of two group LNGCA component time courses differed across groups (p$<0.01$), while the variance of their matched GIFT component time courses did not; two GIFT components differed across groups, while variance of their matched group LNGCA components did not; and one component was significant in both methods. 
To further investigate the differences between the methods, we focused on the three ASD vs TD comparisons with the smallest p-values.  
We report the log-variance of subjects' time series corresponding to the three selected comparisons with the smallest p-values in  \cref{fig:pvalue}, which depicts means +/- 2SE and violin plots of the densities for each group. 
We also plot the corresponding components in \cref{fig:signals}.

In \cref{fig:signals}, we see that component 59 differs greatly between the two methods, component 47 is highly similar, and component 57 has some differences. For components 59 and 47, we only observed group differences in intrinsic engagement using estimates from group LNGCA (\cref{fig:pvalue}). Component 59 is particularly poorly matched across methods (cross method correlation = .26); the group ICA version of this component was labeled as artifact, while the group LNGCA version is more easily identified as part of the DMN. 
The group LNGCA result suggests that component 59 is less intrinsically active in the ASD group than the TD group (\cref{fig:pvalue} bottom right). Extant ASD research  \citep{padmanabhan2017default} has consistently implicated disruptions in intrinsic activity within the DMN, which purportedly contributes to ASD-associated difficulties integrating information about the self in the context of others.   The spatial representation of component 57 is moderately similar between the two methods (cross method correlation = .62) and also appears to be driven by regions of the DMN, but the group LNGCA version is slightly more left-lateralized. Consistent with the group LNGCA result for component 59, the log variances of the subject-specific time courses for component 57 suggest it is less intrinsically engaged in the ASD group compared to the typically developing group, as shown in \cref{fig:pvalue}. 
The group difference we observed was consistent across methods for component 47.  The spatial representations of this component were very similar across group ICA and group LNGCA (correlation=0.91, \cref{fig:signals} middle panel). Component 47, which was labeled as belonging to the frontoparietal network, was less engaged in the ASD group than in the TD group using estimates from both group ICA and group LNGCA (\cref{fig:pvalue} left). This finding is also consistent with existing literature \citep{yerys2019functional} implicating reduced connectivity within the frontoparietal network in children with ASD.

We also examined subject-specific components (Web Appendix \cref{fig:IndividualComps}). The subject specific components include motion-related artifacts characterized by activation near the edge of the brain (first and third rows), as well as other scanner artifacts.

\section{Discussion}
\label{s:discuss}

We propose a method to extract group non-Gaussian components from hundreds of subjects. We demonstrate in simulations that our method can extract low variance features that are discarded using group ICA. Our method involves a first-stage LNGCA for each subject. In this stage, we present a novel test of the number of non-Gaussian components in the presence of  spatially correlated noise that improves upon methods assuming uncorrelated noise, which dramatically overestimate the dimensionality. We apply group LNGCA to an rsfMRI study and discover components that exhibit different levels of activity in children with ASD as compared with typically developing children. 

Group LNGCA divides each subject's components into group, individual, and Gaussian noise components. Current group ICA methods generally assume common group components with subject-specific temporal deviations, including back projection and dual regression (e.g.,  \citealt{erhardt2011comparison}) and hierarchical ICA \citep{guo2013hierarchical}. Our formulation can allow subject-specific deviations to be captured by individual components. These individual components can also correspond to artifacts, and the subject-level LNGCA step appears to separate artifacts from neuronally related components. Lower dimensional PCA in PCA+ICA can mistakenly aggregate features, as examined in \citet{risk2019linear}, although here we have compared LNGCA to PCA+ICA with relatively high dimensional PCA. In some respects, our formulation is more flexible since it can detect components unique to one subject. Our framework is related to Joint and Individual Variation Explained 
\citep{lock2013joint} 
and Common Orthogonal Basis Extraction (COBE) \citep{zhou2015group}. Individual subspaces in rsfMRI improved the prediction of behavior  \citep{kashyap2019individual}. This is also a promising direction for future research in group LNGCA.

In GIFT, there is arguably no clear guidance how to select the number of PCs for both the subject and group level. For the subject-level, it is often selected as many as the computation can afford, but not smaller than the estimated number from the MDL-based method, which assumes Gaussianity. In group LNGCA, our proposed test can be used to determine the number of components to be kept for the subject level.  Determining the number of group components is unresolved in the group ICA literature and is beyond the scope of this work. Methods from JIVE could be examined to determine whether subject-level non-Gaussian components are noisy estimates of group components \citep{lock2013joint, feng2018angle}.

Lastly, our application to resting-state fMRI may be a conservative illustration of the difference between group ICA and group LNGCA. We chose 85 PCs and 85 NGs for group ICA and group LNGCA, respectively, which is higher than suggested by GIFT's criterion, and hence is higher than would be used in many neuroimaging studies. 
Even with this conservative approach, we find important differences in resting-state networks. Most notably, the component 59 showed significantly greater activity in the ASD versus typically developing group in group LNGCA, but these differences were not detected using the matched component from group ICA. We hypothesize that greater dimension reduction in the subject-level analyses would lead to larger differences in the estimated resting-state networks.




\section*{Acknowledgements}

Work was supported in part by grants from Autism Speaks, the National Institute of Mental Health (R01 MH078160, U54 HD079123, K01 MH109766), the National Institute of Neurological Disorders and Stroke (R01 NS048527), the National Institutes of Health/National Center for Research Resources Clinical and Translational Science Award (UL1 TR 000424-06), and the National Institute of Biomedical Imaging and Bioengineering (P54 EB15909). Financial support from the Cornell University Institute of Biotechnology, the New York State Division of Science, Technology and Innovation (NYSTAR), a Xerox PARC Faculty Research Award, National Science Foundation Awards 1455172, 1934985, 1940124, and 1940276, USAID, and Cornell University Atkinson’s Center for a Sustainable Future, is gratefully acknowledged as well.



\bibliographystyle{apalike}

\bibliography{groupngca,references}

\begin{appendix}
		\clearpage
	\subfile{WebSupplement.tex}

\end{appendix}

\end{document}

%% file: WebSupplement.tex
\doublespacing
	\begin{center}
		\Large\textbf{Web-based Supplementary Materials for ``Group linear non-Gaussian component analysis with applications to neuroimaging''}\\
	\end{center}


\section{Simulation details}

\subsection{Experimental setup}
\label{sec:sim_setup}
We used the time courses estimated from group LNGCA of the real data application to estimate a time series model for use in the simulations. For each of the $85$ non-Gaussian components extracted for each of the $342$ subjects, we first estimated the time courses by ordinary least squares, and then fitted the estimated time courses using an AR(1) process. The median of the estimated coefficient is $0.37$ over all component/subject combinations (85*342). We used this AR coefficient to simulate time courses following an AR(1) process in our simulations.

\subsection{Details on dimension estimation}
\label{sec:sim_dim_est}
\begin{figure}
    \centering
    \includegraphics[width=0.8\linewidth]{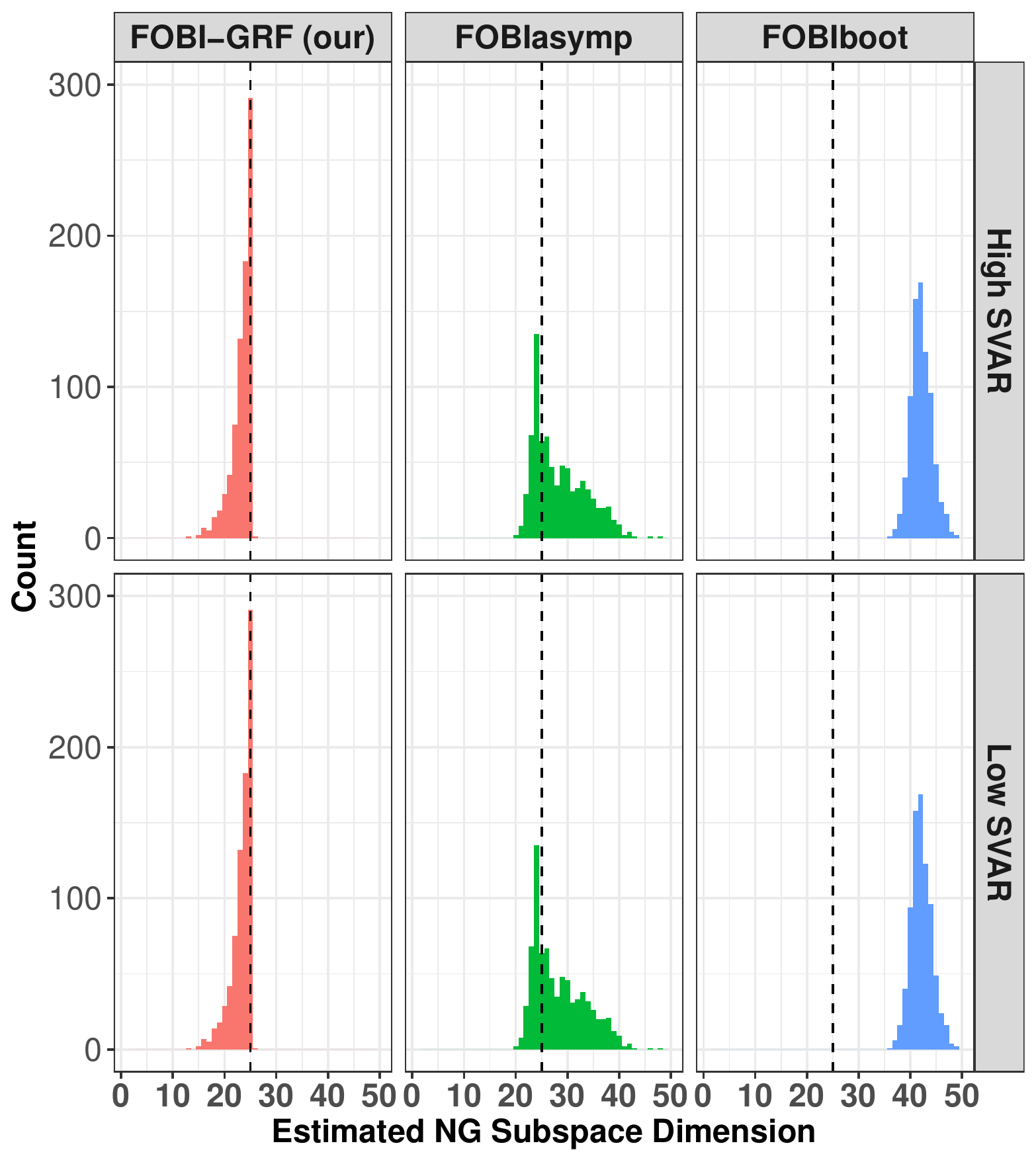}
    \caption{Estimated non-Gaussian subspace dimension across $800$ subjects under high and low SVAR setting. The significance level $\alpha=0.05$. Dashed line indicates true dimension $25$. The most frequently selected dimension using our test, FOBI-GRF, corresponds to the true dimension. Although the test underestimated the dimensions in many simulations, this was due to possibly missing individual components, while it always retained the  group components.}
    \label{fig:dimest_high_low}
\end{figure}

In this section, we provide the results of dimension estimation.
In short, all three methods return very similar but not identical dimension estimation across three SVAR settings, displayed in \cref{fig:dimest_high_low}.
We conjecture the reason is that the variance does not drive the dimension estimation accuracy, but rather the difference in the non-Gaussianity of the non-Gaussian components and the non-Gaussianity achieved by rotating the spatially correlated Gaussian noise.

\subsection{Details on group components extraction}
\label{sec:sim_comp_est}
In this section, we provide the results of group LNGCA and group ICA on group components in simulations when the number of group components is misspecified: $q_G=4$ and $q_G=2$.

\begin{figure}
    \centering
    \includegraphics[width=0.9\linewidth]{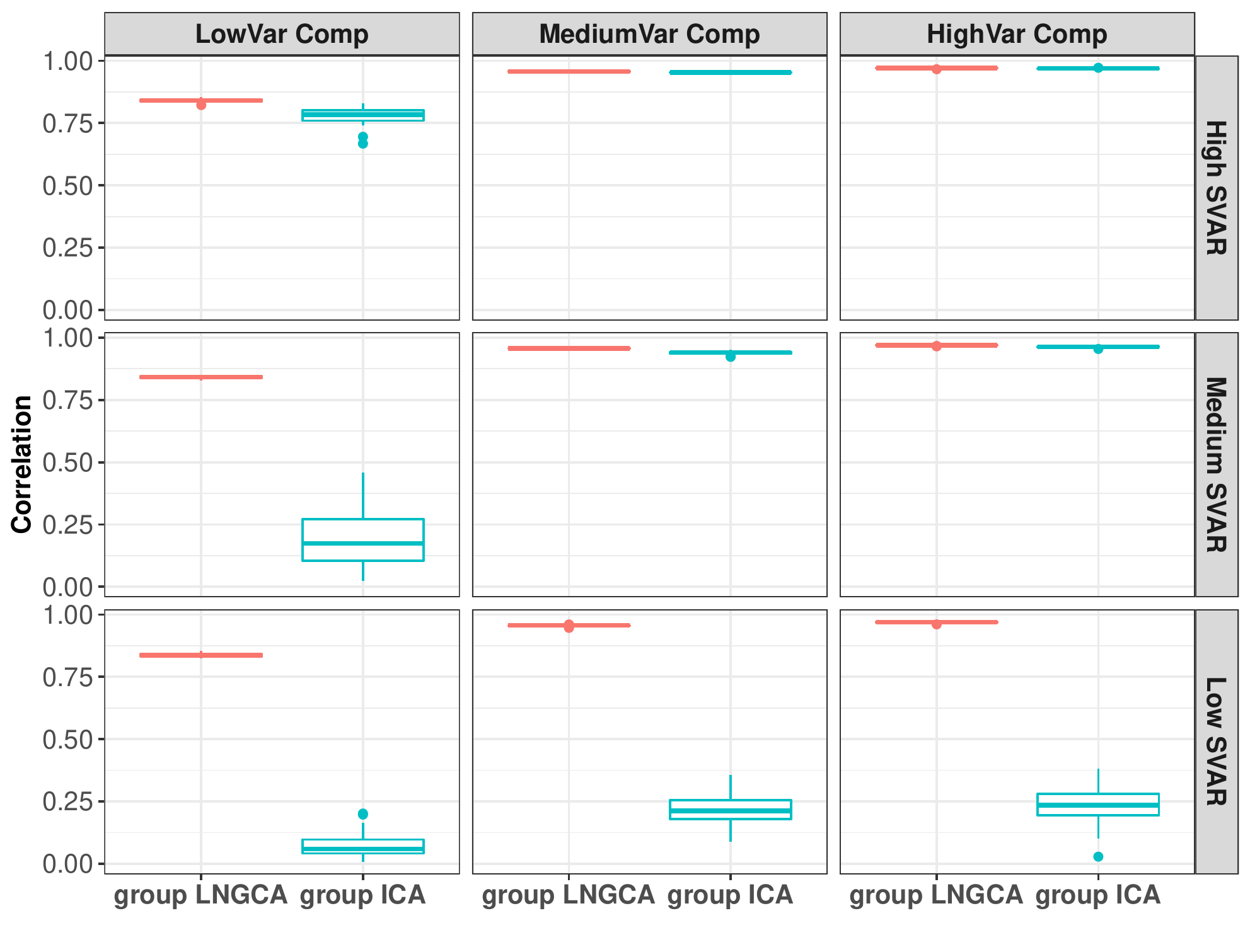}
    \caption{Correlation between three true group components and their matched estimated components when $q_G=4$. The correlation under all settings for group LNGCA concentrates at a high correlation value with vanishing variance.}
    \label{fig:corr_4comp}
\end{figure}

\begin{figure}
    \centering
    \includegraphics[width=0.9\linewidth]{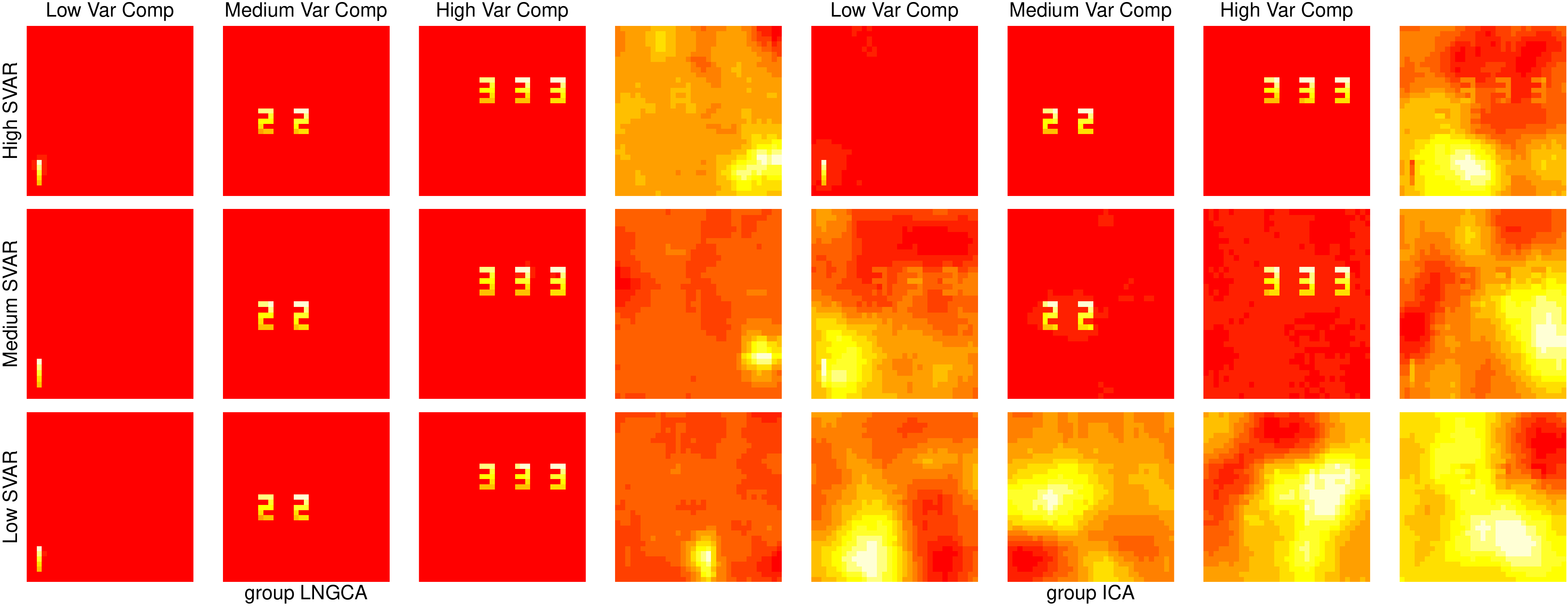}
    \caption{The estimated group components from a representative simulation (median matching error) when $q_G=4$. Left four columns display results from group LNGCA, while right four columns display results from group ICA. For each method, the allocated variance increases from left to right among three signals excluding the last estimated noise signal. Three rows represents high, medium and low SVAR settings respectively, from top to bottom.}
    \label{fig:comp_est_4comp}
\end{figure}

First when $q_G=4$,
both methods perform similar to the case when $q_G=3$:
the three components matched to true group components have similar correlation as when $q_G=3$,
showed in \cref{fig:corr_4comp}.
The estimated signals from the repetition associated with the median matching error (with true signal) in each setting
are depicted in \cref{fig:comp_est_4comp}. 

\begin{figure}
    \centering
    \includegraphics[width=0.9\linewidth]{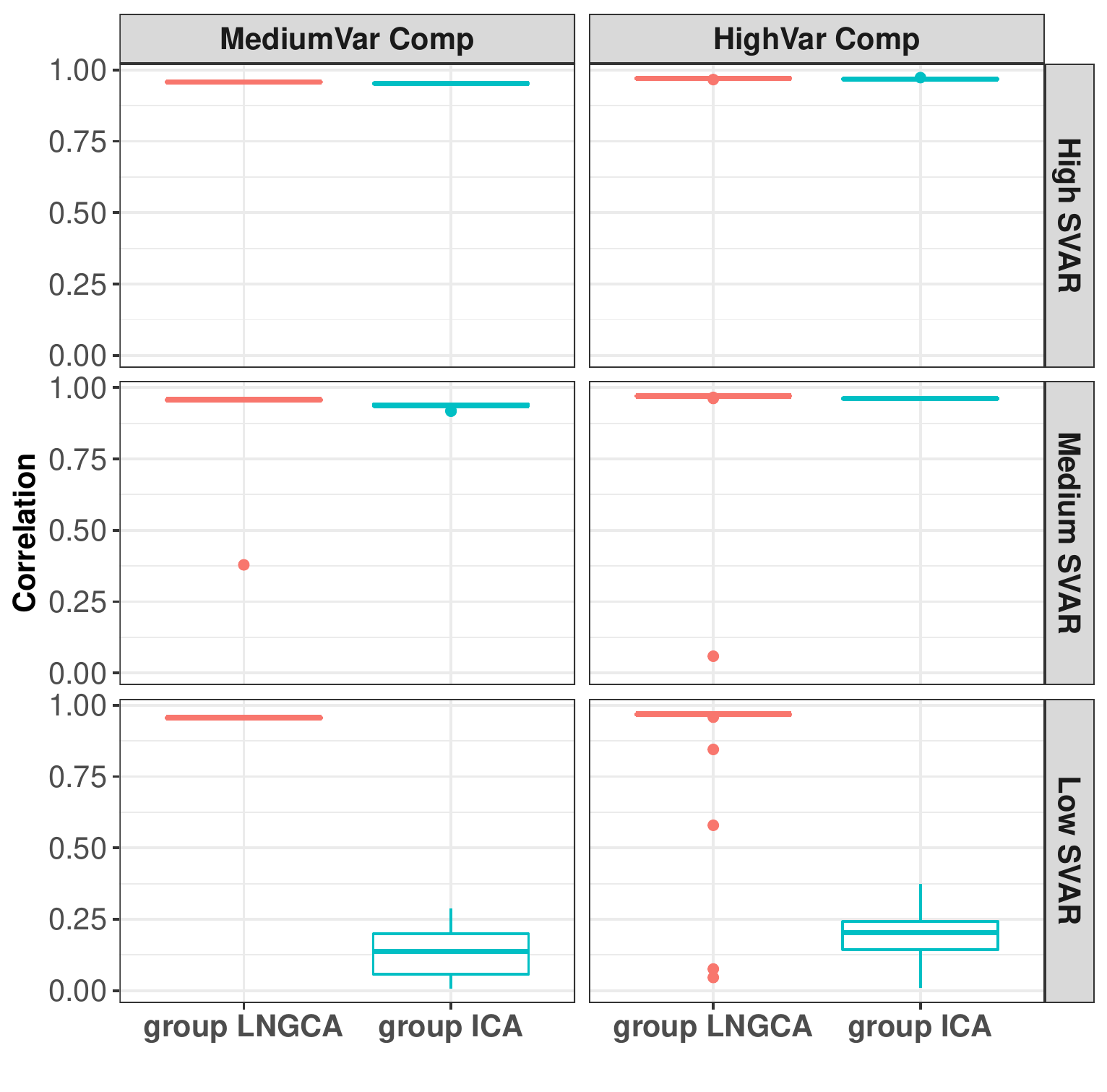}
    \caption{Correlation between two estimated components and their matched true group components when $q_G=2$. The correlation under all settings for group LNGCA concentrates at a high correlation value with vanishing variance.}
    \label{fig:corr_2Comp}
\end{figure}

\begin{figure}
    \centering
    \includegraphics[width=0.9\linewidth]{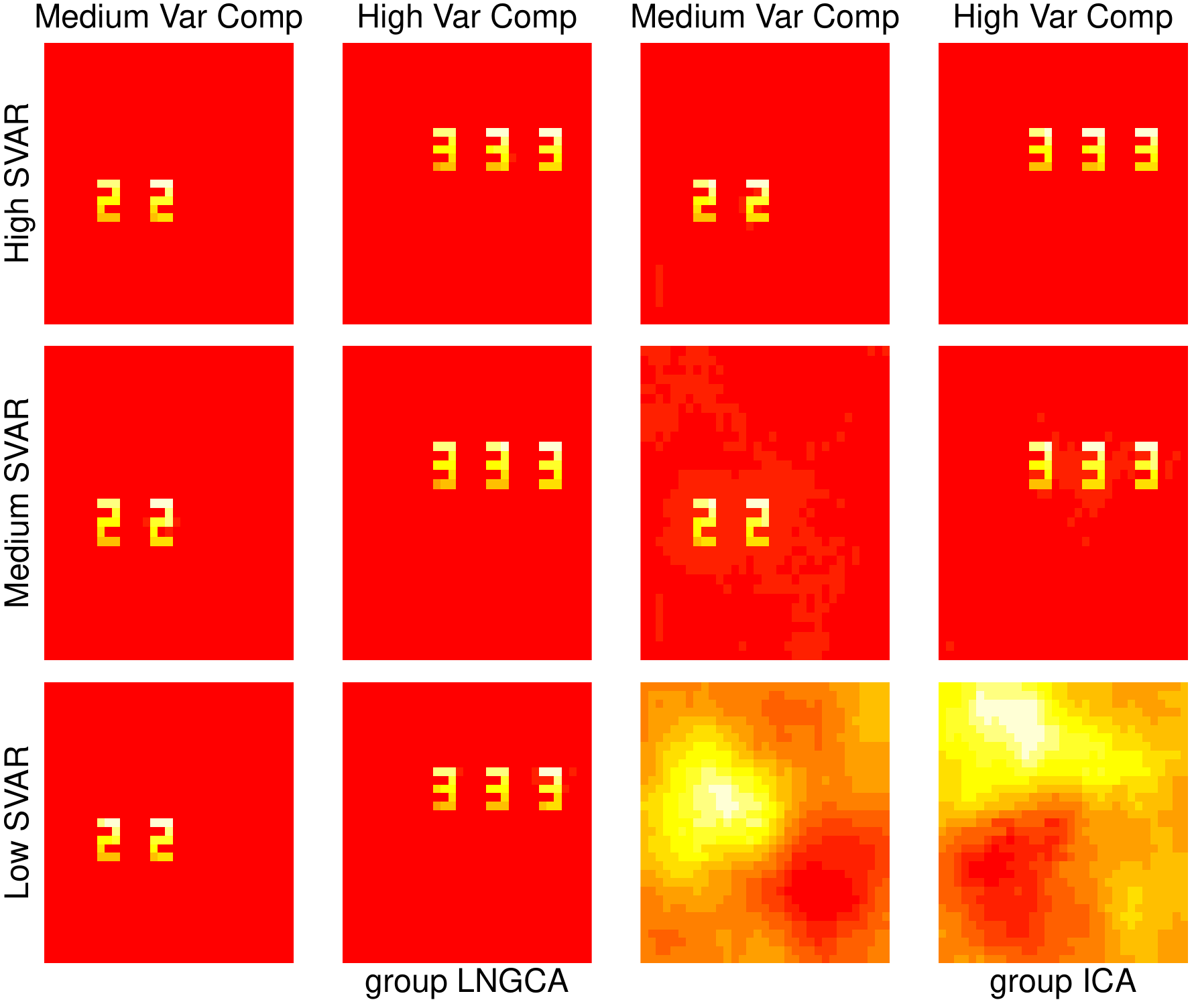}
    \caption{The estimated group components from a representative simulation (median matching error) when $q_G=2$. Left two columns display results from group LNGCA, while right two columns display results from group ICA. For each method, the allocated variance increases from left to right. Three rows represents high, medium and low SVAR settings respectively, from top to bottom.}
    \label{fig:comp_est_2comp}
\end{figure}

When $q_G=2$,
both methods extract two components with higher variance. 
The correlation between true group components and their matched estimation are similar to the corresponding value when $q_G=3$,
showed in \cref{fig:corr_2Comp}.
The estimated signals from the repetition associated with the median matching error (with true signal) in each setting
are depicted in \cref{fig:comp_est_2comp}.   

\section{Details on resting-state fMRI data example}

\subsection{Additional information on the data and preprocessing}
\label{sec:real_data_preprocessing}

All children completed a mock scan to acclimate to the scanning environment. Participants were instructed to relax, fixate on a cross-hair, and remain as still as possible. Functional data were preprocessed using SPM12 and custom MATLAB code (\url{https://github.com/KKI-CNIR/CNIR-fmri_preproc_toolbox}). Rs-fMRI scans were slice-time adjusted using the slice acquired in the middle of the TR as a reference, and rigid body realignment parameters were estimated to adjust for motion. The volume collected in the middle of the scan was non-linearly registered to Montreal Neurological Institute (MNI) space using the MNI EPI template. The estimated rigid body and nonlinear spatial transformations were applied to the functional data in one step, producing 2-mm isotropic voxels in MNI space. Voxel time series were linearly detrended. Data were excluded for between-volume translational movements $>3$-mm or rotational movements $>3$ degrees.  

Group ICA and its PCA steps were applied using GIFT. The second stage PCA was implemented using multi-power iterations \citep{rachakonda2016memory}. 

\subsection{Dimension estimation}
\label{sec:real_data_dim_est}

We applied the NG subspace dimension test in Section 2.3 to six participants. The estimated dimension was 65, 38, 89, 47, 73 and 83, corresponding to participants with 128 time points for first three participants and 156
for last three participants. We report the sequential hypothesis applied, its associated p-values and computation time here for each participant in Table \ref{table:dimtest}. For each participant, the sequential test starts from 85, and the estimated dimension is determined by the two bold test results. 

We also implemented a sequential test with \texttt{FOBIasymp} and the estimated dimensions are 126, 126, 126, 148, 151 and 153 for participants \#1,\ldots,\#6 correspondingly. Such large dimension will not help reduce much computation in practice. It also implies \texttt{FOBIasymp} tends to overestimate the number of non-Gaussian components, as discussed in our simulation. 

\begin{table}
\caption{Details on dimension test of six subjects}
\centering
\begin{tabular}{c|lc}
			\toprule
			  & Sequential Hypothesis Path, presented as dimension(p-value) & Time (mins)  \\
			\midrule
			\#1  & 85(0.02)-43(0.03)-\textbf{64(0.04)}-75(0.11)-70(0.14)-67(0.10)-66(0.06)-\textbf{65(0.07)} & 94    \\
			\#2   & 85(0.72)-43(0.07)-22(0)-33(0.01)-\textbf{38(0.06)}-36(0.04)-\textbf{37(0.03)}& 134   \\
			\#3   & 85(0.01)-106(0.13)-96(0.11)-91(0.11)-\textbf{88(0.04)}-90(0.1)-\textbf{89(0.12)} & 54   \\
			\#4   & 85(0.7)-43(0.02)-64(0.27)-54(0.14)-49(0.06)-\textbf{46(0.03)}-48(0.08)-\textbf{47(0.06)} & 200   \\
			\#5   & 85(0.11)-43(0)-64(0.03)-75(0.08)-70(0.02)-\textbf{73(0.08)}-\textbf{72(0.04)} & 151  \\
			\#6   & 85(0.06)-43(0)-64(0.03)-75(0.04)-80(0.03)-\textbf{83(0.06)}-\textbf{82(0.04)} & 149  \\
			\bottomrule
		\end{tabular}
		\label{table:dimtest}
\end{table}


\subsection{Subject-specific components}
\begin{figure}
\centering
\includegraphics[width=\textwidth]{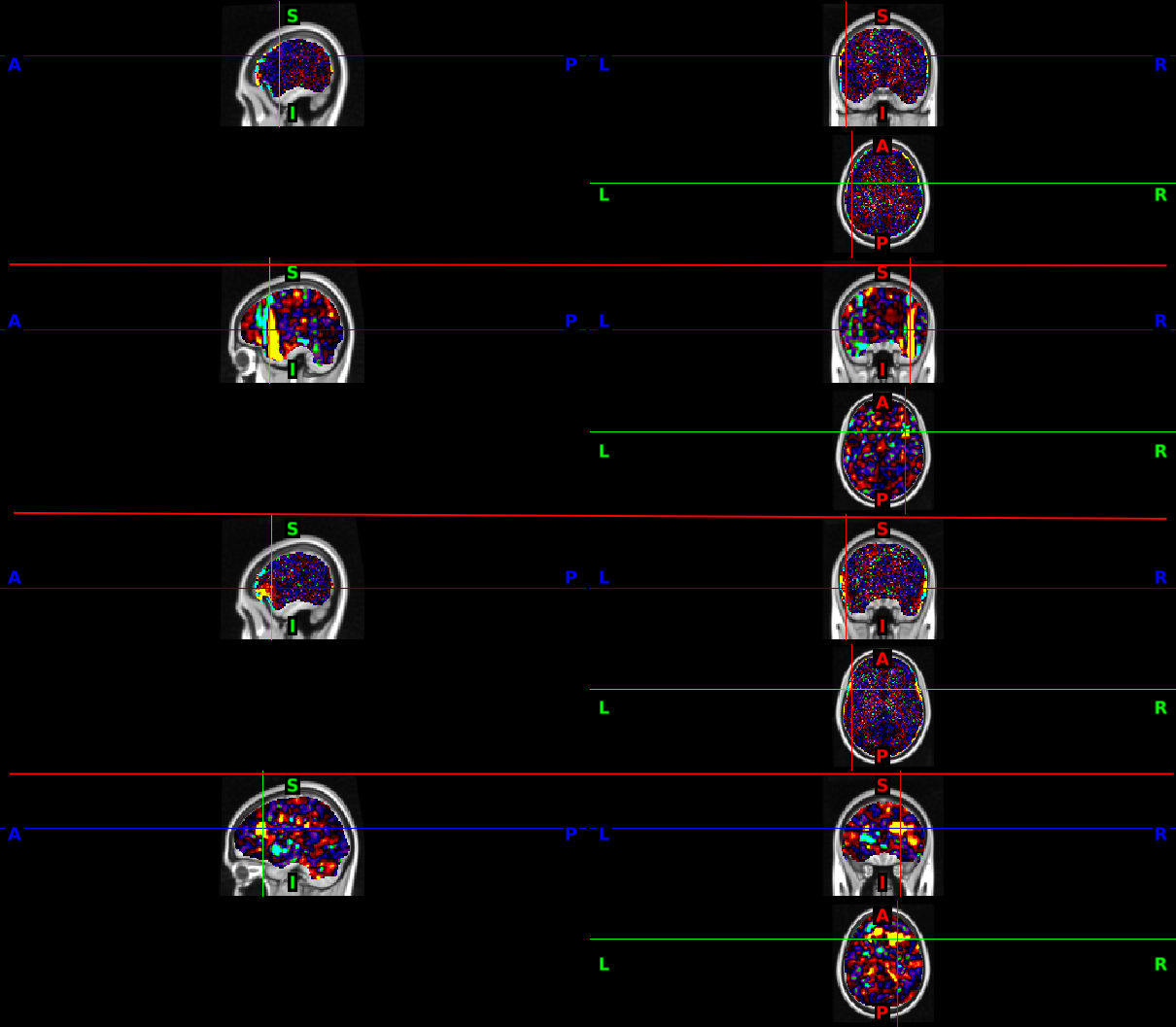}
\caption{Example subject-specific components from four different participants. These components include artifacts. Activation near the brain edge, as in the first and third rows, is often indicative of a motion artifact.}\label{fig:IndividualComps}
\end{figure}

Example subject-specific components are depicted in Figure \ref{fig:IndividualComps}.
